\documentclass[%
reprint,aip,
sd,%
amsmath,amssymb,
]{revtex4-1}

\usepackage{graphicx}
\usepackage{dcolumn}
\usepackage{bm}

\begin{document}
	
\preprint{AIP/123-QED}
	
\title{Coulomb interactions induced perfect spin filtering effect in a quadruple quantum-dot cell}

\author{Maxim Yu.\, Kagan}
\email{kagan@kapitza.ras.ru}
\affiliation{P.L. Kapitza Institute for Physical Problems RAS, 119334 Moscow, Russia}%
\affiliation{National Research University Higher School of Economics, 101000 Moscow, Russia}

\author{Valerii V.\, Val'kov}%
\email{vvv@iph.krasn.ru}
\affiliation{Kirensky Institute of Physics, Federal Research Center KSC SB RAS, 660036 Krasnoyarsk, Russia}%

\author{Sergei V.\, Aksenov}
\email{asv86@iph.krasn.ru}
\affiliation{Kirensky Institute of Physics, Federal Research Center KSC SB RAS, 660036 Krasnoyarsk, Russia}

\date{\today}

\begin{abstract}
	A quadruple quantum-dot (QQD) cell is proposed as a spin filter. The transport properties of the QQD cell were studied in linear response regime on the basis of the equations of motion for retarded Green's functions. The developed approach allowed us to take into account the influence of both intra- and interdot Coulomb interactions on carriers' spin polarization. It was shown that the presence of the insulating bands in the conductance due to the Coulomb correlations results in the emergence of spin-polarized windows (SPWs) in magnetic field leading to the high spin polarization. We demonstrated the SPWs can be effectively manipulated by gate fields and considering the hopping between central dots in both isotropic and anisotropic regimes.
\end{abstract}


\pacs{73.63.Kv, 73.21.La, 73.23.Ra, 81.07.Ta} 
\keywords{Spin filter, Quantum interference, Fano-Feshbach resonance, Coulomb correlations}
\maketitle

\section{\label{sec1}Introduction}
The generation of tunable highly spin polarized current is vital for spintronic applications \cite{prinz-98,wolf-01}. To achieve this aim the variety of systems has been already proposed from semiconductor heterostructures to mesoscopic samples \cite{zutic-04,torio-04}. In the former the electron spins are controlled by the Rashba spin-orbital coupling (SOC) \cite{rashba-60,rashba-84}. The strength of the SOC in turn can be regulated by an electric field perpendicular to 2D electron gas \cite{koga-02}. Along with the SOC in mesoscopic devices, having at least a few Feynman paths, quantum interference in phase-coherent transport regime plays important role \cite{bercioux-05}. In some works it was demonstrated that the interplay between the Aharonov-Bohm (AB) flux \cite{aharonov-59} and the Rashba SOC results in a substantial spin-polarized conductance \cite{sun-05,aharony-08}. However, the experimental implementation of such low dimensional nanosystems, in particular, varying the SOC strength by electric field or penetrating the AB ring with magnetic field seems to be rather difficult.

It is known that the structures having the AB geometry or the networks consisting of quantum dots (QDs) are able to exhibit the Fano-Feshbach resonance \cite{fano-61,feshbach-62} in their transport characteristics as well. As a result, the Zeeman splitting of spin-dependent conductances in the region of such an asymmetric peak leads to the emergence of so-called spin-polarized window (SPW) when there is high probability of tunneling for the electrons with spin $\sigma$ and close-to-zero one for the electrons with spin $\overline{\sigma}$ \cite{wu-04,gong-06,ojeda-09,fu-12}. For the QD-networks previously proposed as spin-filter prototypes in \cite{gong-06,ojeda-09,fu-12} it is highly preferable to have many QDs considering the Coulomb correlations inside each QD but not between them. In this article we will show that the nanosized diamond-shaped cell composed of just four QDs, a quadruple quantum-dot cell (QQD cell), can display perfect spin filtering properties. This behavior is achieved by taking into account both intra- and interdot Coulomb interactions. It is shown that its spin polarization can be effectively manipulated by different kinetic processes in the cell and gate fields.

\section{\label{sec2}The model}
\begin{figure}[tb]
	\includegraphics[width=0.48\textwidth]{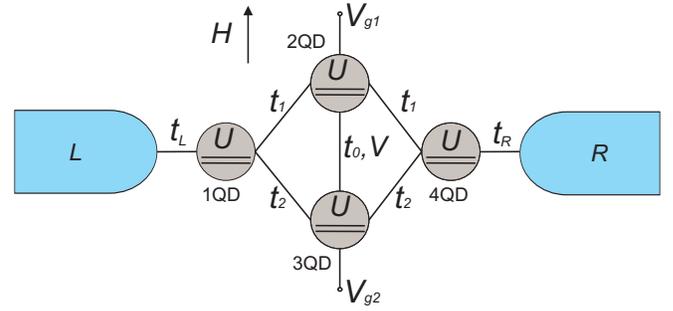}
	\caption{\label{model} The QQD cell between paramagnetic leads.}
\end{figure}
The system under consideration is a QQD cell between paramagnetic contacts in external magnetic field $H$ depicted at figure \ref{model}. It is modeled by the Hamiltonian $\hat{H} = \hat{H}_L + \hat{H}_R + \hat{H}_{D} +\hat{H}_T$. The term $\hat{H}_{L\left(R\right)}$ describes the left (right) lead,
\begin{equation} \label{HLR}
	\hat{H}_{L\left(R\right)}=\sum\limits_{k\sigma}\xi_{k\sigma}
	c^+_{L\left(R\right)k\sigma}c_{L\left(R\right)k\sigma},
\end{equation}
where $c^+_{L\left(R\right)k\sigma}$ is the creation operator in the left (right) lead with quantum number $k$, spin $\sigma$ and spin-dependent energy $\xi_{k\sigma}=\epsilon_{k\sigma}-\mu$; $\mu$ is the chemical potential of the system. The third term describes the QQD cell
\begin{eqnarray} \label{H4l}
	&&\hat{H}_{D} =\sum\limits_{\sigma;j=1}^{4}\xi_{j\sigma}a^+_{j\sigma}a_{j\sigma}+U\sum\limits_{j=1}^{4}n_{j\uparrow}n_{j\downarrow}\nonumber\\
	&&+V\sum\limits_{\sigma\sigma'}n_{2\sigma}n_{3\sigma'}+ \sum\limits_{\sigma}\left[t_{1}\left(a^+_{1\sigma}+a^+_{4\sigma}\right)a_{2\sigma}\right.\\
	&&\left.+t_{2}\left(a^+_{1\sigma}+a^+_{4\sigma}\right)a_{3\sigma}+t_{0}a^{+}_{2\sigma}a_{3\sigma}+h.c.\right],~\nonumber
\end{eqnarray}
where $a_{j\sigma}$ annihilates the electron with spin $\sigma$ and energy $\xi_{j\sigma}=\varepsilon_{j}-\sigma h-\mu$ on the $j$th QD; $h=\mu_B H$ is the Zeeman energy; $\sigma=\pm1$ or $\uparrow,\downarrow$; $t_{i}$ ($i=0,~1,~2$) is a hopping matrix element between the QDs; $U$ is the intensity of the intradot Coulomb interaction; $V$ is the intensity of the interdot Coulomb interaction between the electrons in the $2$nd and $3$rd QDs.

The interaction between the leads and the QQD cell is determined by the last term in the Hamiltonian
\begin{equation} \label{HT}
	\hat{H}_T =t_{L}\sum \limits_{k\sigma}c_{Lk\sigma}^{+} a_{1\sigma} +
	t_{R}\sum \limits_{k\sigma} c_{Rk\sigma}^{+} a_{4\sigma} + h.c.,
\end{equation}
where $t_{L\left(R\right)}$ is a hopping matrix element between the left (right) lead and the $1$st ($4$th) QD.

\section{\label{sec3}Conductance of the QQD cell with Coulomb correlations}
It is convenient to introduce new second quantization operators, $\widehat{\psi}_{\sigma}=\left(a_{1\sigma}~...~a_{4\sigma}\right)^T$, for the QQD cell which allow us to consider the cell effectively as a one-level QD. In this article we present the investigation of the transport properties of the QQD cell in the linear response regime and at low temperatures. This case is correctly described in terms of spin-dependent transmission, $T_{\sigma}$, by the Landauer-Buttiker formula,
\begin{eqnarray}\label{G}
	&&G=-G_{0}\sum\limits_{\sigma}\int\limits_{-\infty}^{+\infty}d\omega \frac{\partial f}{\partial \omega} T_{\sigma}\left(\varepsilon_{D},~\omega\right)=G_{\uparrow}+G_{\downarrow},\nonumber\\ &&T_{\sigma}=\widehat{\Gamma}_{L}\widehat{G}_{\sigma}^{r}\widehat{\Gamma}_{R}\widehat{G}_{\sigma}^{a},~
	\widehat{G}_{\sigma}^{a}=\left(\widehat{G}_{\sigma}^{r}\right)^+,
\end{eqnarray}
where $G_{0}=e^2/h$ is the conductance quantum; $f\left(\omega\right)$ is the Fermi distribution function. The matrix $\widehat{\Gamma}_{L\left(R\right)}$ describing the coupling between the left (right) lead and the device is supposed to be spin- and frequency-independent since the paramagnetic leads are treated at the wide-band limit. The first (last) diagonal element of the matrix, $\Gamma_{L\left(R\right)}=\pi t_{L\left(R\right)}^2\rho_{L\left(R\right)}$ ($\rho_{L\left(R\right)}$ is the constant density of states of the leads), is the only nonzero one. 

In order to find the components of the retarded matrix Green's functions of the cell taking into account the intra- and interdot Coulomb interactions we solved the equations of motion for its components, $G_{i\sigma j\sigma'}^{r}\left(\omega\right)=\langle\langle a_{i\sigma} | a_{j\sigma'}^{+} \rangle\rangle$,
\begin{equation} \label{eqG1}
	z\langle\langle a_{i\sigma} | a_{j\sigma'}^{+} \rangle\rangle=\left\langle\left\{a_{i\sigma},~a_{j\sigma'}^{+}\right\}\right\rangle+
	\langle\langle \left[a_{i\sigma},~\hat{H}\right] | a_{j\sigma'}^{+} \rangle\rangle,
\end{equation}
where $z=\omega+i\delta$. In general the presence of nonlinear terms in $\hat{H}_{D}$ gives rise to infinite set of equations which includes the hierarchy of the all-order Green's functions, such that $\langle\langle n_{j\overline{\sigma}}a_{j\sigma} | a_{j\sigma}^{+} \rangle\rangle$, $\langle\langle n_{i\sigma}a_{j\sigma} | a_{j\sigma}^{+} \rangle\rangle$, $\langle\langle n_{i\overline{\sigma}}a_{j\sigma} | a_{j\sigma}^{+} \rangle\rangle$ and so on. To truncate this set and get closed one we employ the procedure used by You and Zheng \cite{you-99a,you-99b}. This decoupling scheme allows to consider the Coulomb correlations beyond the Hartree-Fock approximation. In the same time spin-flip processes leading, in particular, to the Kondo physics are neglected \cite{lacroix-81}. The final system of equations involves the first, second and third order Green's functions (for details see \cite{kagan-16}). The retarded Green's functions, the occupation numbers and correlators are calculated self-consistently using additionally the kinetic equations,
\begin{eqnarray}
	&&\langle n_{i\sigma}\rangle=-\frac{1}{\pi}\int d\omega f\left(\omega\right) Im\left[G^{r}_{i\sigma,i\sigma}\left(\omega\right)\right],\label{kineq}\\
	&&\langle a_{i\sigma}^{+}a_{j\sigma}\rangle=-\frac{1}{\pi}\int d\omega f\left(\omega\right) Im\left[G^{r}_{j\sigma,i\sigma}\left(\omega\right)\right],~i\neq j.\nonumber
\end{eqnarray}
In this article we focus on spin filtering properties of the QQD cell in the presence of the Coulomb interactions. The corresponding spin polarization coefficient is
\begin{equation}\label{P}
	P=\frac{G_{\uparrow}-G_{\downarrow}}{G_{\uparrow}+G_{\downarrow}}.
\end{equation}
\begin{figure*}[htbp]
	\includegraphics*[width=0.5\textwidth]{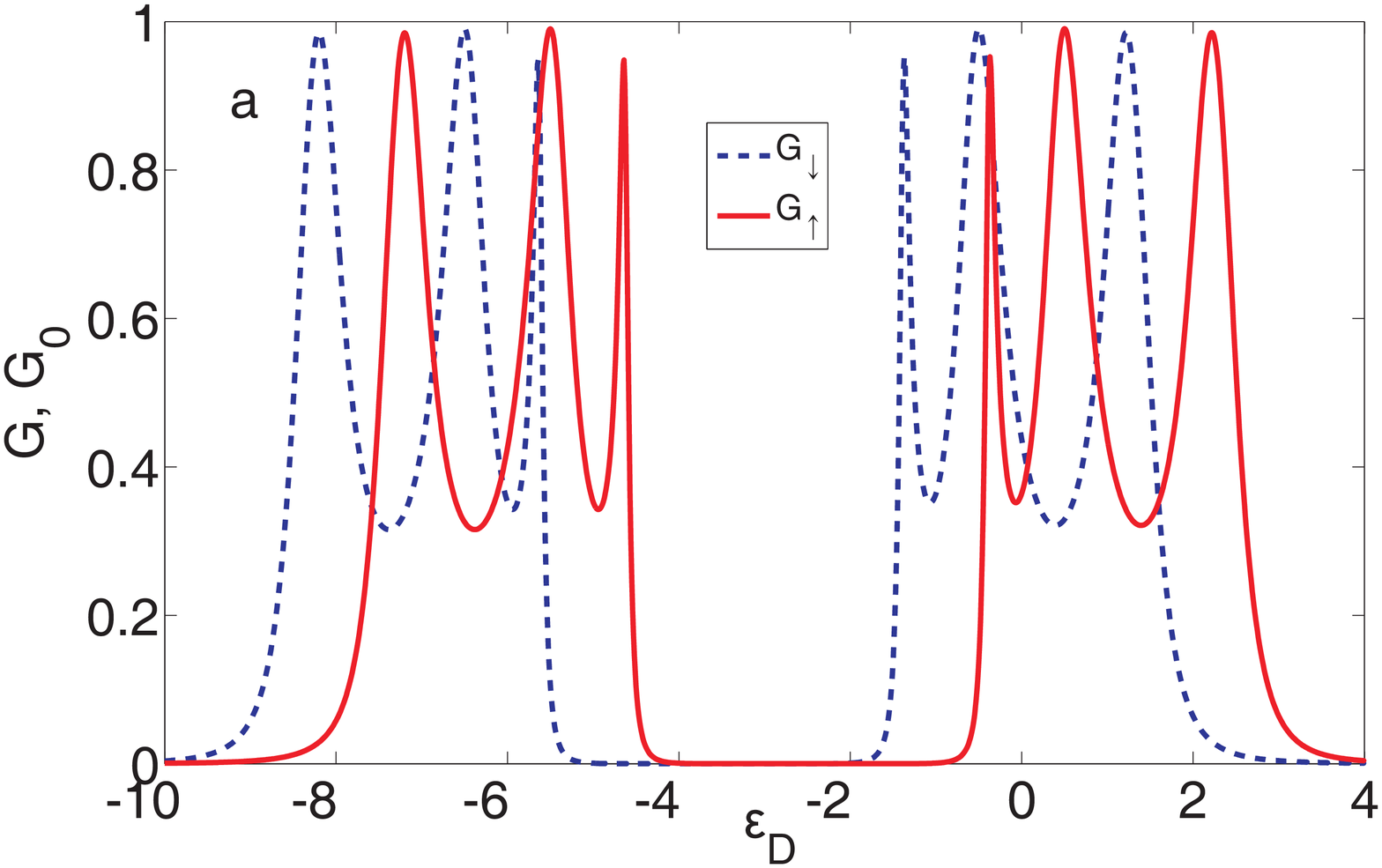}
	\includegraphics*[width=0.49\textwidth]{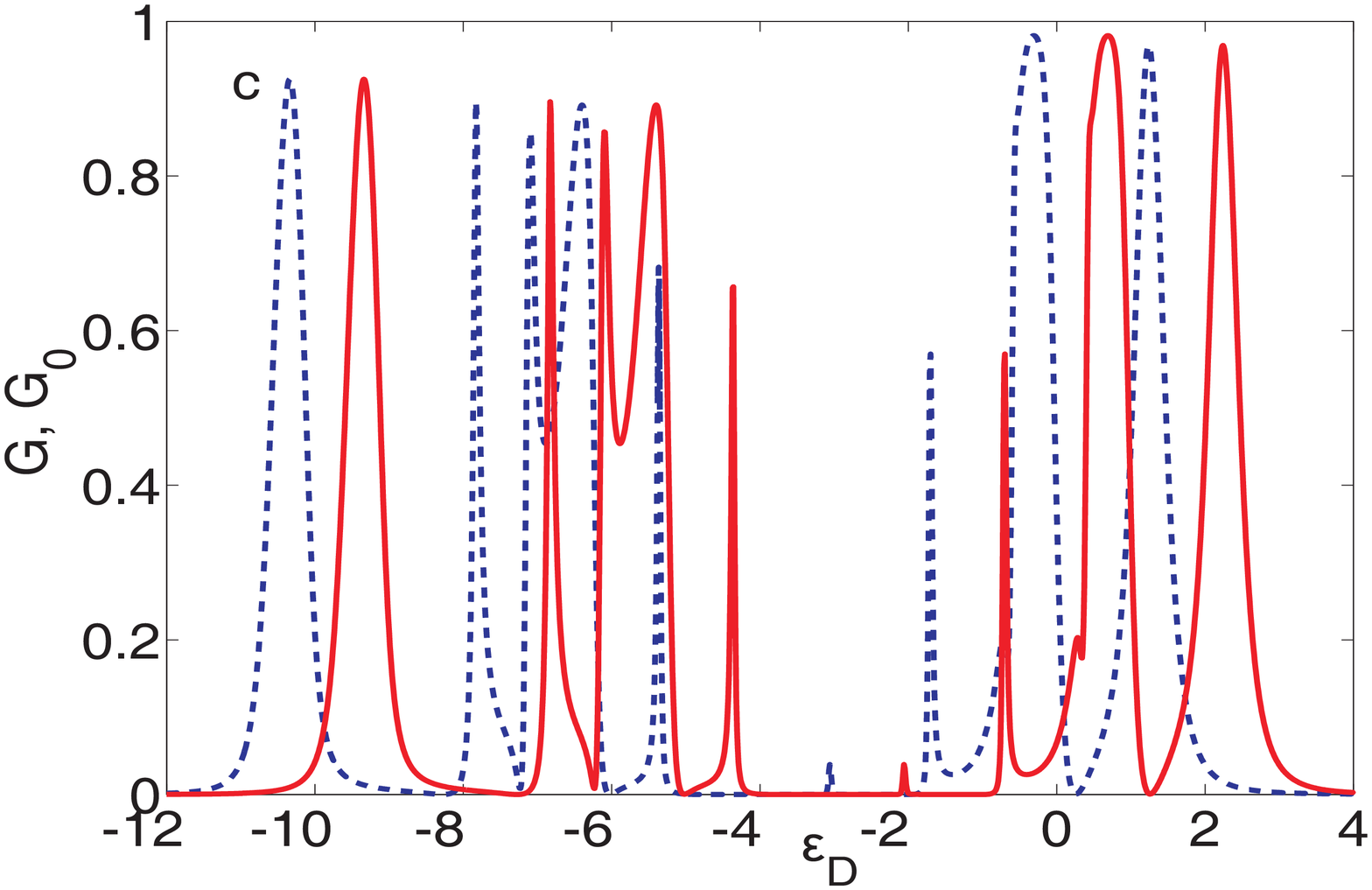}
	\includegraphics*[width=0.49\textwidth]{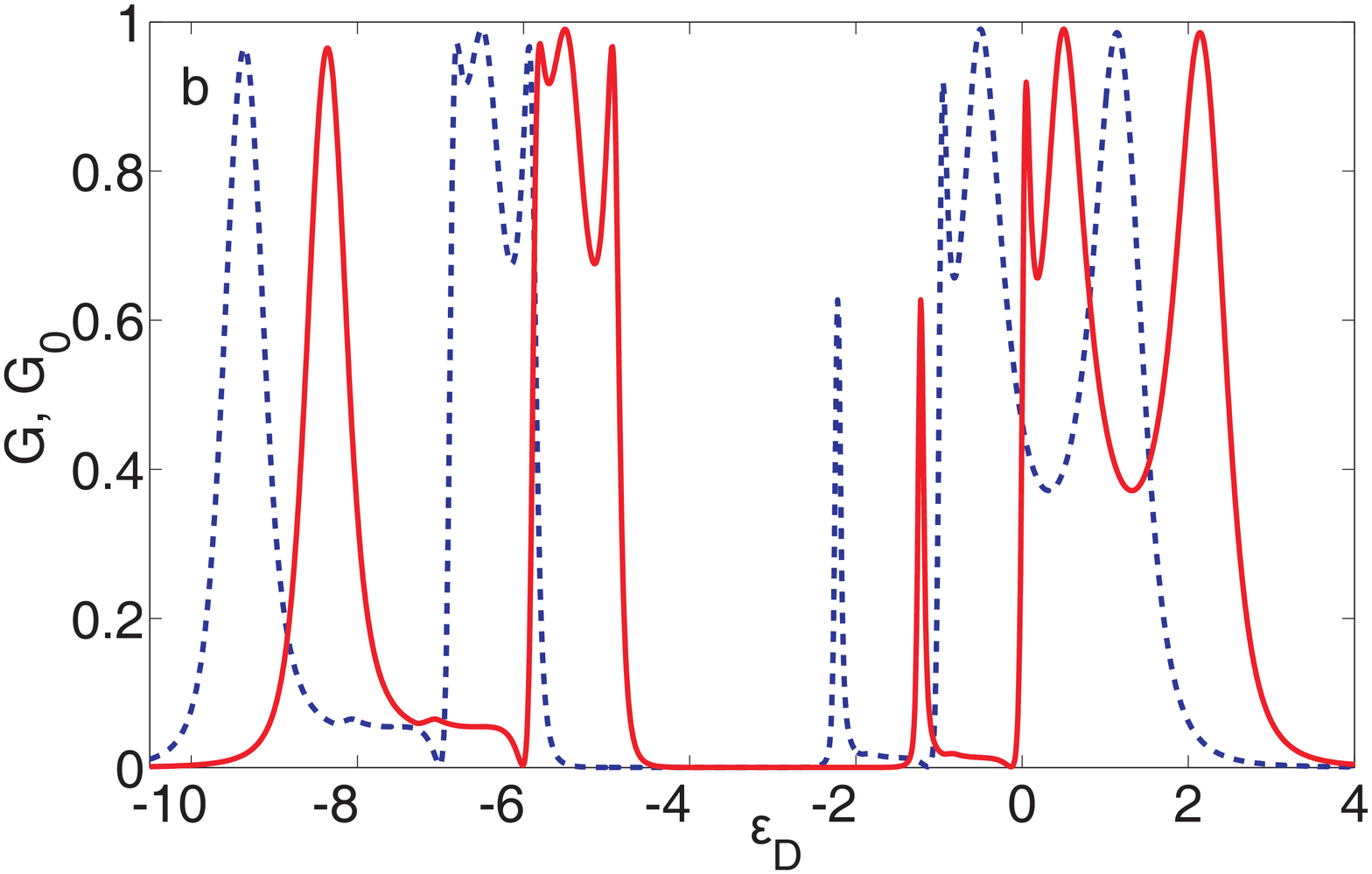}
	\includegraphics*[width=0.5\textwidth]{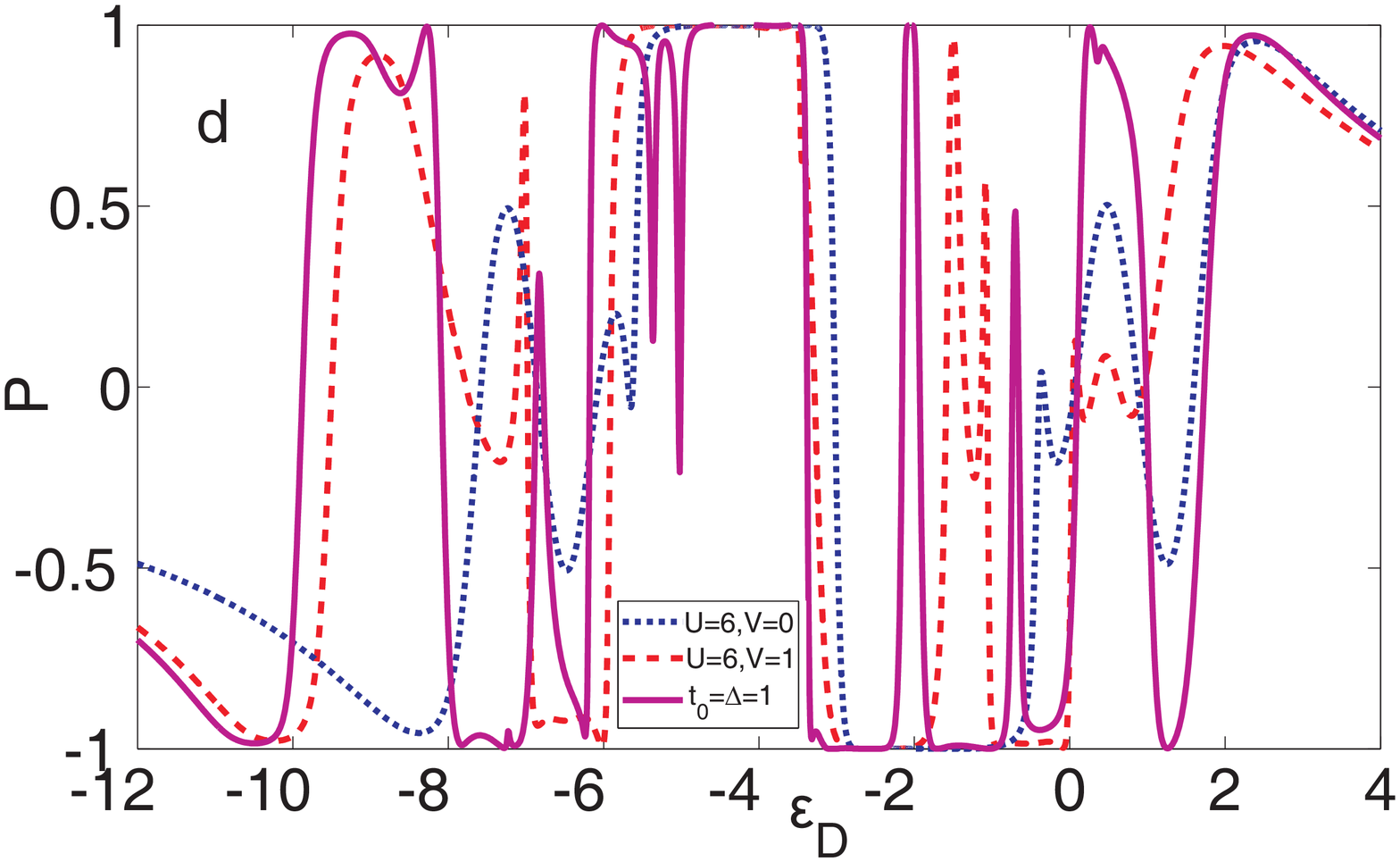}
	\caption{\label{2} The spin-up, $G_{\uparrow}$, and spin-down, $G_{\downarrow}$, conductance of the isotropic QQD cell: (a) $U=6$, $V=t_0=\Delta=0$; (b) $U=6$, $V=U/6$, $t_0=\Delta=0$; (c) $U=6$,~$V=U/6$, $t_0=\Delta=1$; (d) the spin polarization. Other parameters: $k_{B}T=0.01$, $h=0.5$.}
\end{figure*}
We study here a symmetrical transport situation in strong coupling regime, $t=t_{1}$, and use $\Gamma_{L}=\Gamma_{R}=t$ in energy units.

\section{\label{sec4}Isotropic QQD}
Initially we analyze the transport properties of the isotropic QQD cell when $t_1=t_2$ and $\varepsilon_{j}=\varepsilon_D$. As it is clearly seen from the figures \ref{2}a-c the application of the magnetic field on the cell causes the separation of the spin-up (solid curves) and spin-down (dashed curves) conductances. Consequently, at some gate and magnetic fields perfect spin filtering can be achieved when the conductance of electrons with the spin $\sigma$ is close to zero but the conductance for opposite spin $\overline{\sigma}$ is significant and even approaches unity (in units of quantum conductance $G_0$). In this desirable for spintronic applications case, according to \eqref{P}, $P=\pm1$. When only the intradot Coulomb interactions are taken into account we get two triple-peak structures (TPSs) due to the electron-hole symmetry (fig. \ref{2}a) \cite{chen-94}. The number of the resonances in each TPS corresponds to the number of the QDs in top and bottom paths for electronic waves. Between two TPSs there is the insulating band with steep edges. As a result, the step-like feature emerges in $P$ at gate fields $\varepsilon_D\approx-5$ $\div$ $-1$ (see dotted curve at fig. \ref{2}d). At upper and lower plateaus of $P$ we receive perfect spin polarization of carriers with opposite sign due to the appearance of SPWs in the conductance. The most interesting situation takes place if the interdot Coulomb interaction between the electrons of the 2nd and 3rd QDs is considered along with the intradot correlations. In this regime additional wide low-conductance band is induced by the Fano-Feshbach resonance after half filling (see e.g. $G_{\uparrow}$ at gate fields $\varepsilon_D\approx-8$ $\div$ $-6$ at fig. \ref{2}b) \cite{kagan-16}. The Zeeman shift of the spin-dependent conductance, such that the Fano antiresonance of $G_{\downarrow}$ coincides with the corresponding peaks of $G_{\uparrow}$, leads to new step-like feature in $P$ at $\varepsilon_D\approx-7$ $\div$ $-5$ (see dashed line at fig. \ref{2}d). Integrally, three zones with high spin-up polarization ($P>0.5$) and four zones with high spin-down one ($P<-0.5$) are generated by the Coulomb correlations in the cell. The unequal number of the zones for $sgn\left(P\right)=\pm1$ and their difference before and after half filling ($\varepsilon_D\approx-3$ for $G_{\uparrow}$) is explained by the breaking of the electron-hole symmetry due to the interdot repulsion \cite{chen-94}. Moreover, we considered two more ways to manipulate $P$ by creating additional SPWs, namely by means of the hopping between central QDs, $t_{0}$, and making the energy levels of the central QDs nonidentical by using gate fields, $\xi_{2\sigma}=\xi_{3\sigma}+2\Delta$. At figure \ref{2}c the total effect of both factors is demonstrated. It is clearly seen that there are more the spin-up and spin-down SPWs due to new Fano-Feshbach resonances. Thus, the spin polarization is consecutively switched between the conducting channels with $P\approx\pm1$ as $\varepsilon_D$ is swept (see solid line at fig. \ref{2}d).

\section{\label{sec5}Anisotropic QQD}
\begin{figure*}[htbp]
	\includegraphics*[width=0.49\textwidth]{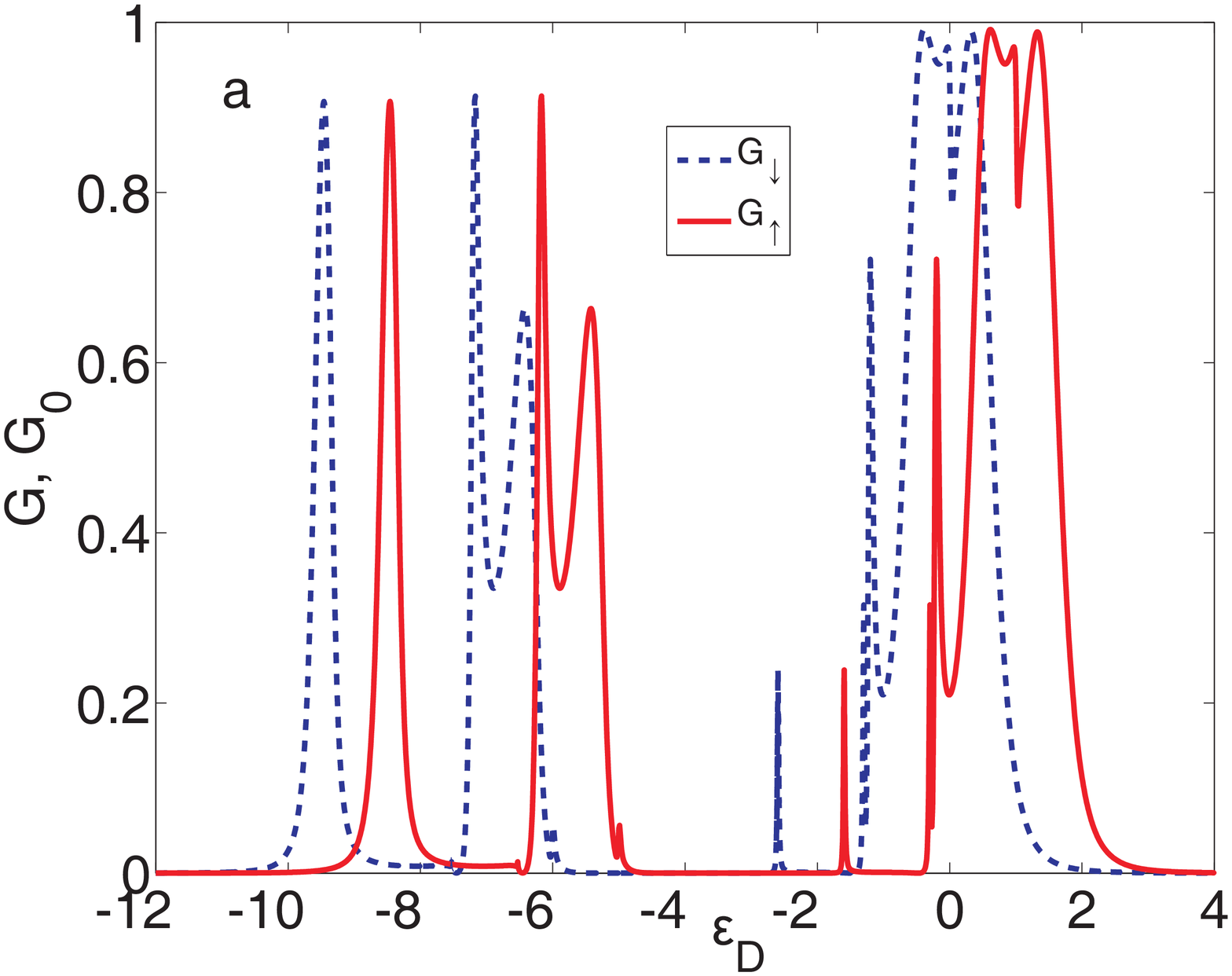}
	\includegraphics*[width=0.49\textwidth]{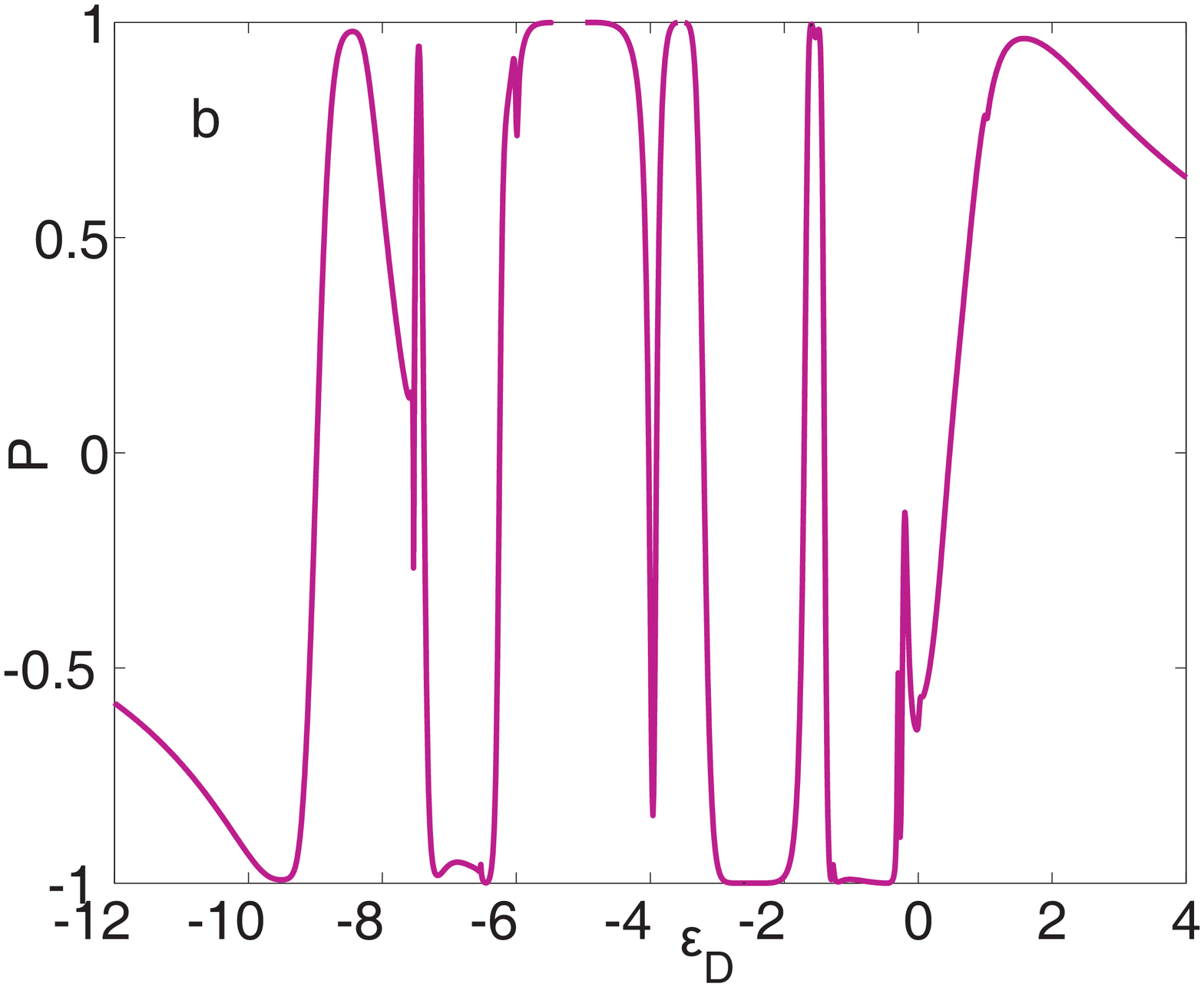}
	\caption{\label{3} (a) The spin-dependent conductance of the anisotropic QQD cell for parameters of figure \ref{2}c, $t_2=0.1$, $t_0=0.2$, $\Delta=0.5$. (b) The spin polarization.}
\end{figure*}
From experimental point of view, it is natural to consider an anisotropic QQD cell where the transfer integrals differ from each other. In particular, we suppose here the top path is more transparent than the bottom one, i.e. $t_1 \gg t_2,~t_0$. Additionally, it is worth to remark that such a system is to some extent analogous to the two-band Hubbard systems with one narrow band, especially exhibiting electron polaron effect \cite{kagan-11a,kagan-11b}. The decreasing of $t_2$ and $t_0$ leads to the suppression of narrow conductance peaks in comparison with the isotropic case (see figures \ref{3}a and \ref{2}c respectively). Consequently, six explicit SPWs (three for each spin projection) are formed by the Zeeman shift and the spin polarization at figure \ref{3}b has a set of the zones with high $\mid P \mid$. Therefore, it is strictly shown that the QQD cell in both isotropic and anisotropic configuration can be used as a perfect spin filter. This feature is based on the presence of the intra- and interdot Coulomb interactions in the structure.

\section{\label{sec6}Conclusion}
In this article the spin filtering properties of the QQD cell in the presence of the external magnetic field have been analyzed. Using the equation-of-motion technique for retarded Green's functions we showed that intra- and interdot Coulomb interactions of the carriers in the cell lead to the appearance of SPWs in the conductance. They correspond to the zones of high spin polarization of the current. The width and quantity of the SPWs can be controlled by gate fields and the ratio of the transfer integrals in the cell. The switching between the perfectly spin-polarized transport channels was demonstrated in both isotropic and anisotropic QQD cell.

\begin{acknowledgements}
	We acknowledge fruitful discussions with P.I. Arseyev, N.S. Maslova, V.N. Mantsevich and R.Sh. Ikhsanov. This work was supported by the Comprehensive programme SB RAS no. 0358-2015-0007, the RFBR, projects nos. 15-02-03082, 15-42-04372, 16-42-243056, 16-42-242036 and partly by Government of Krasnoyarsk Region. M.Yu. K. thanks the Program of Basic Research of the National Research University Higher School of Economics for support.
\end{acknowledgements}

\bibliography{KM_VV_AS_APL}

\end{document}